\documentclass[twocolumn,showpacs,preprintnumbers,amsmath,aps,amssymb]{revtex4}
\usepackage{graphicx}% Include figure files
\usepackage{dcolumn}% Align table columns on decimal point
\usepackage{bm}% bold math

\begin{document}

\title{Is noncommutativity related with the smallness of $\Lambda$?.}
 \author{E. Mena}
  \altaffiliation[Permanent address:]{
Centro Universitario de la Cienega,\\
Ave. Universidad 1115 Edif. B, C.P. 47820
Ocotl\'an,\\ Jalisco, M\'exico}
\email{emena@cuci.udg.mx} 
\author{O. Obreg\'on}
\email{octavio@fisica.ugto.mx}
\author{M. Sabido}
\email{msabido@fisica.ugto.mx}
\affiliation{ Instituto de F\'{\i}sica de la Universidad de Guanajuato,\\
 A.P. E-143, C.P. 37150, Le\'on, Guanajuato, M\'exico\\
 }%

\date{\today}% It is always \today, today,
\begin{abstract}
In this letter we study the effects of a noncommutative minisuperspace, including matter degrees of freedom  on a FRW universe with cosmological constant. In this setting the vacuum energy density  can be calculated to be of the same order as the observed energy density of the universe.
 \end{abstract}
 \maketitle
%\section{Introduction}
The cosmological constant problem has been addressed by means of different approaches for several years and still today remains as one of the central issues of not only modern day cosmology but also particle physics \cite{polchinski}. The remarkable discovery and confirmation of the acceleration of the universe is usually attributed to a small, non-vanishing $\Lambda$. Unfortunately there is no known mechanism that guarantees a zero or nearly zero value for $\Lambda$ in a stable or metastable vacuum. This has been complicated by the fact that is associated energy density has a similar value as that of the energy of present day matter.
These and other questions connected to the cosmological constant can be casted as three fundamental problems: why is the cosmological constant so small?, why it is not zero? and why is it comparable to the matter energy-density (cosmic coincidence)?. The apparent impossibility of addressing these questions has lead to speculate on the necessity of an anthropic principle \cite{garriaga}, but several alternatives have been suggested (see \cite{stefan} and references there in).

In this paper we will focus on the first question (as nicely put by Polchinski \cite{polchinski} ``it is hard enough"), also known as the ``old cosmological constant problem". In a more precise manner, why is the effective cosmological constant $\Lambda_{eff}$ so close to zero. The different contributions to the vacuum energy density, from ordinary particle physics should give a value for $\langle\rho\rangle$ of order $M_p^4$,  which should be canceled by the bare value of $\Lambda$. This cancellation  has to be better than $10^{-121}$ if we compare the zero-point energy of a scalar field, using the Planck scale as a cut-off, to the experimental value of $\langle\rho_{obs}\rangle\approx10^{-47}(GeV)^4$.  This incredible degree of fine tuning,  suggests that we are missing important physics. Then its likely  that the correct way to interpret the tiny value of the cosmological constant by conventional quantum field theory is not the whole story. This has lead many authors to try different approaches to dark energy.

One interesting possibility  lies in the old idea of noncommutative space-time \cite{snyder}. This idea has seen renewed interest as a consequence  of the developments in M-theory and string theory \cite{connes,ref3}. But still one can ask of the relevance of noncommutativity in the cosmological scenario or its relevance at the early universe. { In quantum gravity the measurement of length is limited to distances grater that $l_p$, because in order to locate a particle we would need an energy grater then $M_p$. The corresponding gravitational field will have an horizon $R=\frac{2GM_p}{c^2}=2L_p$. Therefore a minimal size should exist for quanta  of space and time configurations. This can be written as an uncertainty relationship between the space-time coordinates, that would arise from a non zero commutation relation between the coordinate operators. From this, one would expect that at very early times of the universe non trivial effects of noncommutativity be present.
In order to consider noncommutativity among the gravitational variables one would, in principle, need a noncommutative theory of gravity \cite{ncsdg,ncsdg2}. Because we are interested in the very early universe we would need more than this, we need a noncommutative quantum theory of gravity. This seems today an impossible task, as we still do not even have a final quantum theory of gravity \cite{qg}, much less  a noncommutative one.
To try to understand the possible influence of noncommutativity at the beginning of our universe we will make use of a previously proposed formalism \cite{ncqc}. The main ingredients of this proposal are: quantum cosmology in the minisuperspace approach, whose variables are the 3-metric components in a finite configuration space. This formalism has the advantage that the inclusion of matter is straight-forward. By considering these models one freezes out degrees of freedom and the canonical quantization of these minisuperspace models gives the Wheeler-DeWitt equation (WDW). 
On the other hand more general analysis  suggest that  conditions can be found to justify the minisupespace approach and presume the behavior of the wave function as fundamental \cite{halliwell}. If one considers string theory, general relativity and consequently the WDW equation, corresponds to the s-wave approximation \cite{susskind}. Secondly, space-time noncommutativity has as a consequence that the fields do not commute. In a more specific manner this is  due to the Moyal product \cite{connes,ref3}. To introduce this elements we take into account the  procedure to generalize usual quantum mechanics to its noncommutative version \cite{gamboa}. Having the WDW equation to describe the quantum evolution of the universe and being the ``coordinates" of these models the fields, it was assumed that the variables  do not commute. Then, an effective noncommutativity was defined in the minisuperspace from which the quantum evolution of the Kantowski-Sachs cosmology was studied \cite{ncqc}.}  In particular in the last few years there have been several attempts to study the possible effects of 
noncommutativity in the cosmological scenario. In \cite{kallosh} it is argued that there is a possible relation between the 4D cosmological constant and the noncommutative parameter of the compactified space in string theory. In \cite{brand} the effects of noncommutativity during inflation are explored, but noncommutativity is only incorporated in the scalar field neglecting the gravitational sector.
 We will consider a model in our 4D space-time and noncommutativity in both the gravitational and matter sectors. 
In our approach we don't intend to explain the origin of the cosmological constant,   however,  we will show that by means of minisuperspace noncommutativity a small cosmological constant arises. It alleviates the discrepancy between the calculated and observed vacuum energy density. Actually in \cite{polchinski}, it is suggested that the cosmological constant problem might require a form of UV/IR mixing, and this is a natural feature of noncommutative quantum field theory \cite{nekrasov}. By combining these different ideas a simple toy model can be constructed. 

Our starting point is a very conventional action,  composed by the gravity sector, the matter sector  and cosmological constant $\Lambda$ 
\begin{equation}
S_{tot}= S_g +S_{m} =\int  d^4x \sqrt{-g} \left[ R -\Lambda+ 
\mathcal{L}_m
  \right] .
\label{lagra}
\end{equation}
This simple model of a time independent cosmological constant is completely consistent with all the available data. Although, some dynamics might be responsible for dark energy, the sensitivity of current observations yields no evidence of evolving dark energy. This model can be considered a solid model for the evolution of the universe \cite{stefan} (except of course, for the high degree of fine tuning).

For the homogeneous, flat and isotropic Friedmann-Robertson-Walker (FRW) metric
\begin{equation}
ds^2= -N^2dt^2 + e^{2\alpha(t)}\left[dr^2
+r^2 (d\vartheta^2 + sin^2 \vartheta d\varphi^2) \right],
\label{frw}
\end{equation}
we define $a(t)=e^{\alpha(t)}$ as the scale factor and $N(t)$ is the lapse function. 
In order to find the dynamics of the universe, we simply solve Einstein equations. We may also proceed by a minisuperspace formalism through hamiltonian dynamics. If we prefer a semiclassical approximation to the WDW equation can be used, of course the method used to find the classical solutions is irrelevant as they are all equivalent. In particular we choose the gauge $N(t)=e^{3\alpha}$  and  as $\mathcal{L}_m$ only the kinetic term for a scalar field $\phi$ with initial momentum $P_{\phi_0}$ (this choice of matter content is motivated since it  is the simplest one that gives the appropriate density perturbations, that are the seeds for structure formation). We are interested in  calculating the vacuum energy an need then the scalar field. 

The solution for the scale factor is  given by
\begin{equation}
% \phi(t)&=&\phi_0-P_{\phi_0}t, \\
 \alpha(t)=\frac{1}{6}\ln{ \left( \frac{P_{\phi_0}^2}{4\Lambda}\right)}
+ \frac{1}{3}\ln{\left(\textrm{sech}\left[\frac{\sqrt{3}}{2}P_{\phi_0}(t-t_0)\right]\right)}.
\label{alpha}
\end{equation}
Although, this equation may seem a little awkward, it describes the appropriate dynamics  in the gauge we have chosen. That means that the clock we are using to measure time is different that the usual one   in order to get the results in cosmological time we would need to set $N(t)=1$. We have, however, the freedom to choose $N(t)$  and assumed this particular choice of gauge because the effects we want to show can be seen  more clearly.

%%%%%%%%%%%%%%%%%%%%%%%%%%%%%%%%%%%%%%%%%%%%%%%%%%%
As mentioned at the beginning of this letter, we want to apply  {\it noncommutativity} to this model, unfortunately  to use a noncommutative theory of gravity is a very difficult ordeal.
{ For this we will follow \cite{ncqc}, one starts by calculating the Hamiltonian from Eq.(\ref{lagra}) together with the FRW metric. This gives a Hamiltonian as a function of the minisuperspace variables $\alpha$, $\phi$ and their canonical momenta. After canonical quantization, this variables are promoted to operators as in usual quantum mechanics and finally arriving to the WDW equation. Following \cite{gamboa,ncqc}, the introduction of noncommutativity is achieved by 
 a noncommutative deformation 
 of the minisuperspace variables is assumed}
\begin{equation}
[\alpha,\phi]=i\theta,
\label{ncms}
\end{equation}
this can be seen as an effective noncommutativity that could arise from a fundamental noncommutative theory of gravity. For example, if we start with the Lagrangian derived in \cite{ncsdg} (this is a higher order Lagrangian and is expanded in the usual noncommutative parameter), the noncommutative fields are a consequence of noncommutativity among the coordinates \cite{ncsdg,Jurco:2001rq} and then the minisuperspace variables would inherit  some effective noncommutativity. This we assume to be encoded in (\ref{ncms}), otherwise we would have a very complicated Hamiltonian for the higher order Lagrangian.

This effective noncommutativity
can be formulated in terms of product of functions of the minisuperspace variables, with the Moyal product of functions.
We know from noncommutative quantum mechanics \cite{gamboa} that the modified  commutator for the   operator associated to the coordinates can be returned to the original commutative  relations if we introduce the following change of variables $
\alpha\to \alpha+\frac{\theta}{2}P_\phi $ and $\phi\to \phi-\frac{\theta}{2}P_\alpha$. It is also easy to show \cite{gamboa}, that the efects of the Moyal star product are reflected in the WDW equation only through a shift in the potential $V(\alpha,\phi)\star\Psi(\alpha,\phi)=V(\alpha+\frac{\theta}{2}P_\phi,\phi-\frac{\theta}{2}P_\alpha)\Psi(\alpha,\phi)$. Applying this to  the WDW equation corresponding to action (\ref{lagra}) we can write
\begin{equation}
\left[ -\frac{1}{24}\frac{\partial^2}{\partial\alpha^2}+\frac{1}{2}\frac{\partial^2}{\partial\phi^2}+2 \Lambda e^{6(\alpha-i\frac{\theta}{2}\frac{\partial}{\partial \phi})} \right]\Psi(\alpha,\phi)=0.
\end{equation}
this is then the noncommutative WDW equation (NCWDW) and its solutions give the quantum description of the noncommutative universe. In \cite{ncqc} the authors arrive to a similar equation, but they are interested only in the quantum epoch of the Kantowski-Sachs universe, this is done by solving the NCWDW equation and plotting the probability density after forming a wave packet.
Our model  will exhibit essentially the same behavior; for particular values of $\theta$ different maxima on the probability density appear, there is one absolute maximum which corresponds to the most probable state of the universe. There are also several other maxima that correspond to different {\it ``universes"}  that can be considered as stable states. This idea is reminiscent of the string theory landscape \cite{polchinski,landscape}, where many stable vacua are present, each vacuum is a local minimum of the superpotential and corresponds to a possible universe. In our case we have several probable universes and each corresponds to a local maximum of the probability density.

We have introduced noncommutativity at the quantum level at the beginning of the universe,  when it was at a quantum state from which it evolved to the classical FRW universe with cosmological constant. If we want to analyze any imprint at  our epoch of the noncommutativity present at the birth of the universe, we should derive from the NCWDW equation the classical evolution. 
In order to achieve this
we  find the temporal evolution of our noncommutative cosmology by a WKB type procedure. For this we propose that the noncommutative wave function has the form $
\Psi_{NC}(\alpha,\phi)\approx e^{i(S_{NC1}(\alpha)+S_{NC2}(\phi))}$, we also take the limit in which the second derivatives of $S_{NC}$ are smaller that the first derivative squared
$\vert S''_{NC}\vert<<\vert S'_{NC}\vert^2$. With these approximations together with the identification $P_{\alpha_{NC}}=\frac{\partial(S_{NC1})}{\partial \alpha}$ and $P_{\phi_{NC}}=\frac{\partial(S_{NC2})}{\partial \phi}$  we find the time dependent solutions for $\alpha_{nc}$
\begin{eqnarray}
\alpha_{nc}(t)&=&\frac{1}{6}\ln{ \left( \frac{P_{\phi_0}^2}{4\Lambda e^{-3\theta P_{\phi_0}}}\right)}\nonumber\\
&+& \frac{1}{3}\ln{\left(\textrm{sech}\left[\frac{\sqrt{3}}{2}P_{\phi_0}(t-t_0)\right]\right)}.
\label{alphanc}
\end{eqnarray}
Comparing this result with the commutative one, we find that the classical evolution of the commutative and noncommutative universes is remarkably the same. From this we can be confident that the phenomenology described by the commutative model can also be explained by the noncommutative model. 
By comparing solutions (\ref{alpha}) and (\ref{alphanc}) we can establish the relationship
\begin{equation}
\Lambda_{nc}=\Lambda e^{-3\theta P_{\phi_0}},
\label{lambda}
\end{equation} 
so the expansion of the universe described by either the commutative  or the noncommutative model is the same and the difference is the value of the cosmological constant. 
This is a very suggestive result, which implies that if we consider a noncommutative universe, the standard value of the cosmological constant is significantly reduced eliminating the necessity of the high degree of fine tuning. On the other hand, one can argue that this simple model can be modified adding other kind of matter for latter stages of the universe to explain other cosmological observations. However, this can be done as in the commutative model, because a modified cosmological constant is the only remnant of the noncommutative quantum universe.

{ As stated at the beginning the problem of the smallness of $\Lambda$ actually means that the rate of $\langle\rho_{obs}\rangle$  to $\langle\rho_{vac}\rangle$ calculated from ordinary particle physics is of order $10^{-121}$.
With this in mind and given the behavior of $\Lambda_{nc}$ we attempt to find the value of the vacuum energy density $\langle\rho_{vac}\rangle_{nc}$ in our noncommutative minisuperspace model}. To calculate the vacuum energy one starts with the energy momentum relationship, write down the Fourier transform of the fields and integrate to a cut-off scale. { Even though, as mentioned, we have not made use of a particular noncommutative theory of gravity \cite{ncsdg,ncsdg2}, our procedure allow us} 
from equations (\ref{alpha}), (\ref{alphanc}) and the definition of the scale factor to establish the relationship between the commutative and noncommutative scale factors $a_{nc}(t)=e^{\theta/2 P_{\phi_0}} a(t)$. In this manner we can define a kind of noncommutative metric
\begin{equation}
g^{(nc)}_{\mu\nu}=diag(e^{3\theta P_{\phi_0}}g_{00},e^{\theta P_{\phi_0}}g_{ij}).
\end{equation}
Calculations should now be performed only with this metric,
which in the linear limit gives the noncommutative equivalent to the Minkowski metric. In order to calculate the vacuum energy density, we must sum the zero point energies of quantum fields in our modified Minkowski metric.  This  is done as in the commutative case but yields a different coefficient which comes from the deformed metric 
\begin{equation}
\langle\rho_{vac}\rangle_{nc}\approx e^{-\theta P_{\phi_0}}k^4_{max},
\label{rho}
\end{equation}
where $k_{max}$ is the fundamental cut-off scale.
One may be tempted  to use different cut-off energies, i.e. grand unification scale or the QCD scale. Because noncommutativity is assumed at the quantum regime of the universe it is expected to be present at Planck's length, then it makes sense to take $k_{max}\approx M_p$.
 As already stated, current observations  put the energy density at a value $\langle\rho_{obs}\rangle\approx 10^{-47}$ and should be of the same order of magnitude as the vacuum energy density. If we consider that the universe is described by the noncommutative model, then we must analyze the ratio between the observed energy density and the vacuum energy density calculated in the noncommutative formalism, this gives the relationship 
%From equations (\ref{lambda}) and (\ref{rho}), we get
\begin{equation}
\frac{\langle\rho_{obs}\rangle}{\langle\rho_{vac}\rangle_{nc}}=e^{\theta P_{\phi_0}}\frac{\langle\rho_{obs}\rangle}{k^{4}_{max}},
\label{razon}
\end{equation}
we note that the ratio of the observed energy density to the cut-off scale is regulated by the exponential $e^{\theta P_{\phi_0}}$. Considering the usual huge discrepancy  of order $10^{-121}$ on the calculated and observed densities, a value of $\theta P_{\phi_0}\approx 240$ can easily suppress it. So, the usual quantum field theory calculation of the vacuum energy density is correct and gives the expected value in the noncommuative universe. Still, the fact  that we need the appropriate initial conditions of the universe remains. Fortunately the effects of the minisuperspace noncommutativity are only reflected through a modified cosmological constant (\ref{lambda}). The type of matter introduced gets dissolved very rapidly as the universe expands leaving its imprint at most on the energy density perturbations.
One should note that the product of $\theta$ and $P_{\phi_0}$ is the relevant parameter. { We have assumed that this product is positive, because one would expect that the introduction of noncommutativity reduces the available physical states (see for example \cite{ncbh}). The reduction of states (that is related to the vacuum energy) has its origin in the introduction of a new fundamental scale of order $\sqrt{\theta}$ that arises from the uncertainty principle $\Delta \alpha\Delta\phi\ge\frac{\theta}{2}$ associated to Eq.(\ref{ncms}). Having a discrete space-time the number of states is diminished in comparison with the continuos commutative space-time}. The effective $\theta$ parameter we have introduced plays its role at the beginning of the universe and then leaves an imprint. 
 
It is worthwhile stressing that this {\bf toy} model shows the possible dramatic influence that noncommutativity could have in the evolution of our present Universe. The old cosmological constant problem has been addressed, by showing that the value of $\Lambda$ can drastically change to an appropriate small value. The fine tuning problem can be softened by considering now the ratio of the observed energy density to the calculated noncommutative vacuum energy Eq.(\ref{razon}). That vacuum energy density  drastically diminishes can be due to the ``discretness" of the noncommutative minisuperspace coordinates. 
The kind of noncommutativity considered here, can in principle be extended to other more realistic cosmological and gravitational models.  Even thought this is only a toy model, it provides a novel way of reasoning these problems. In particular in relation to the central problem of dark energy.

%%%%

\section*{Acknowledgments}
This work was partially supported by CONACYT grants  51306 and 51306,
 and PROMEP grants UGTO-CA-3 and PROMEP-PTC-085.

\end{document}